# Magnetic domain wall propagation in a submicron spin-valve stripe: influence of the pinned layer

J. Briones, F. Montaigne, D. Lacour, M. Hehn
*LPM, Nancy-University, CNRS, BP 329, 54506 Vandoeuvre lès Nancy, France*

M.J. Carey, J.R. Childress
*Hitachi San Jose Research Center, 3403 Yerba Buena Rd, San Jose, California 95135, USA*

**Abstract :**

*The propagation of a domain wall in a submicron ferromagnetic spin-valve stripe is investigated using giant magnetoresistance. A notch in the stripe efficiently traps an injected wall stopping the domain propagation. The authors show that the magnetic field at which the wall is depinned displays a stochastic nature. Moreover, the depinning statistics are significantly different for head-to-head and tail-to-tail domain walls. This is attributed to the dipolar field generated in the vicinity of the notch by the pinned layer of the spin-valve.*

Contact author : montaigne@lpm.u-nancy.fr

In many ways, the control of the position and the propagation of magnetic domain walls (DW) is a key enabler to study various fundamental effects, and a necessary step to investigate depinning mechanisms, the torque exerted by a spin-polarized current on a DW or to design magnetic logic devices [1]. Handling DW in nano-wires appears to be a key milestone towards future spintronic devices [2]. Therefore, various geometries have been considered [3-12] to pin the DW at specific locations. The detection of the DW position has been done either using direct imaging techniques [9-11,13,14] or using position-dependent resistance devices such as giant magnetoresistance (GMR) spin-valves wires [3,4,8,12,13]. In the latter case, a pinned magnetic layer is needed as the reference magnetic layer of the spin-valve. However, its direct influence on DW experiments has not been quantitatively addressed up to now.

In this letter, we highlight the effects of the reference layer on the injection and depinning mechanisms. Most importantly, we show that changing the magnetization configurations of the spin-valve from parallel to antiparallel or from antiparallel to parallel is not equivalent. A fine statistical analysis combined with micromagnetic calculations shows that the presence of the reference layer does not simply induce an homogeneous bias field in the free layer.

The spin-valve was grown by magnetron sputtering on a glass substrate with a structure (in nanometers) Ta(3) / Cu(2) / IrMn(6) / $Co_{65}Fe_{35}$(2.5) / Cu(3) / $Co_{65}Fe_{35}$(4 ) / $Ni_{86}Fe_{14}$(15) / Ru(6). After a thermal anneal, the pinned layer exhibits a exchange bias field of 1.2 kOe and the free layer has a reversal field of 10 Oe. The GMR equals 3%. The spin-valve has been patterned by electron beam lithography and Ar ion beam etching in order to create and manipulate a magnetic DW in a submicron wire 500 nm wide and 20 µm long. A nucleation pad is present on one side of the wire and a notch is positioned along the wire. The pad is designed to inject a DW in the wire at low magnetic fields. Macroscopic Ti/Au leads have been added by a UV lithography process for electric transport measurements. Figure 1.a shows a scanning electron microscopy (SEM) picture of a completed device. GMR measurements allow us to locate precisely the DW between the electrical contacts. AC measurements have been performed using a conventional lock-in 2-contacts technique with an applied current of 10 µA.

In order to suppress any resistance drift due to temperature or variation of contact resistance during the measurements, we use a normalized resistance: 0% corresponding to the parallel state (P) and 100% to the antiparallel state (AP) of the spin valve. Figure 2 represents 4 different cycles obtained by sweeping the field from 300 to -300 Oe (and back) at a rate of 5 Oe/s. Starting from an antiparallel configuration of magnetizations, the AP→P propagation of a domain wall from the injection pad to the notch is associated with a drop in resistance. Magnetic Force Microscopy (MFM) and transport measurements in different geometries (not shown in this paper) revealed that the magnetization in the pad reverses at fields below 30 Oe. But the propagation of this reversed domain is limited by the nucleation pad neck as shown by the MFM image on figure 1(b). Then, this domain is further injected in the wire at the so-called injection field and propagates to the notch. The notch acting as a pinning site stops its propagation (see MFM picture on figure 1(c)). Further increase of the field is necessary to depin the DW. At the so-called depinning field, the normalized resistance of the device drops to 0% corresponding to a fully parallel state. Sweeping the field backwards results in similar phenomena for P→AP propagation. A domain wall is nucleated in the pad at positive fields and propagates up to the notch (as indicated by the change of resistance whom magnitude equals the previous one). At the depinning field, the domain wall propagates to the stripe end and a complete antiparallel state corresponding to 100% of normalized resistance is reached.

Figure 2 clearly shows that this cycle is not fully reproducible. While the domain wall is trapped at the same position (same intermediate resistance), the exact value of the injection and depinning fields varies from experiment to experiment. Such a stochastic behavior has previously been reported [12]. In order to characterize precisely the observed field distributions, more than one thousand GMR loops similar to Fig. 2 have been recorded. Only cycles presenting a proper pinning of the domain wall associated to two well defined transitions are considered [15]. The inset of figure 3 shows such normalized GMR curves measured for P→AP propagation. The stochastic natures of the injection and depinning of the domain wall appears clearly. Therefore, in the following discussion, we only consider the cycles resulting in proper DW trapping. To visualize the statistical data, histograms of observed injection and depinning fields values are represented in figure 3(a) and 3(b), respectively. The statistical distributions are rather complex with different well defined peaks indicating that the involved mechanisms are much more complex than a unique thermally activated process. This complexity might arise from different types of DW. In this range of thicknesses, vortex and transverse domain walls are stable. Moreover, vortex domain walls can vary in chirality and direction of the core; transverse walls can have two orientations. The overall shape and the exact position of trapping may also change. For a single magnetic configuration, the occurrence of different types of depinning mechanisms cannot be excluded [16]. For each magnetic configuration a different depinning field distribution is then expected and the complexity of the measured distribution reflects the variety of possible magnetic configurations. Both the injection and depinning fields present the same kind of complex distribution which is easily understandable since the injection field do not correspond to a nucleation but to the depining of the DW from the narrowing of the pad. It must be noted that no correlation is observed between the injection field and depinning field of the same cycle. This is an indication of the changing nature of the wall during its propagation through the wire. Such a change has already been observed during spin-torque induced DW motion [17,18]

Figure 3 represents the field distributions obtained for both the P→AP and AP→P propagations. Surprisingly these distributions differ significantly. Asymmetry of domain wall propagation has already been reported [6,12,19] for different directions of propagation. But in our case, there is no change in domain wall propagation direction. The domain wall always propagates from the pad towards the notch. The only difference is that for the AP→P propagation a domain parallel to the pinned layer propagates to the detriment of an antiparallel one. A head-to-head DW separates the two domains. Oppositely for P→AP a tail-to-tail DW drives the free layer switching. Due to the time-reversal symmetry, absolutely no difference should exist between the propagation of a head-to-head and tail-to-tail DW's. The only element breaking the symmetry is the presence of the pinned layer whose magnetization is not reversed during the field sweep. For standard Cu thicknesses, there is always, in a continuous film, a finite coupling between the free and pinned magnetic layers of a spin-valve. This coupling is either magnetostatic (Néel or orange-peel coupling) or electronic (RKKY). In our case, the net coupling is positive (favor a parallel alignment of magnetizations) and results in a small shift of 3 Oe in the free layer hysteresis loop. As the pinned layer is not reversed, this coupling acts as a constant effective field on the free layer and could induce a simple shift in the field distributions. Our observations (Fig.3), however, indicate a much more complex behavior not consistent with such a simple picture. The significant difference in the distributions must therefore be due to more complex effects, such as the inhomogenous coupling arising at the edges of the nanostructure. This influence has already been reported in similar micron-sized elements [20, 21]. This coupling is localized where the magnetization of the pinned layer is not parallel to the structure edges, i.e., at the circular side of the pad, at the narrowing of the pad, and at the notch. Let us therefore consider these possible effects in turn. The stray field generated by the pinned layer within the pad can influence the nucleation process and consequently affect the nature of the injected domain wall. This could explain the difference in distribution of the injection field. But as the nature of the domain wall evolves during its propagation, and because the injection and depinning fields are not correlated, this cannot account for the change in distribution of the depinning field. On the other hand, the stray fields present at the narrowing of the pad and around the notch can indeed influence respectively the injection and depinning fields.

In order to verify this hypothesis, the dipolar field generated by the pinned layer must be evaluated. To do this, the magnetic configuration in the pinned layer has been calculated using the OOMMF micromagnetic software [22] considering a 2.5 nm thick Co layer. The effect of antiferromagnetic pinning is modeled by the application of a uniform 1200 Oe field. The idealized geometry reproduced in figure 4 mimics the shape of a real notch. The dipolar field created by this magnetic configuration in the pinned layer is then averaged over the 19 nm thickness corresponding to the free layer. The amplitude of the effective dipolar field acting on the free layer is represented in figure 4(a). *In silico* modeling shows that significant dipolar fields (several tens of Oersted) are present at the notch vicinity, affecting its "pinning potential". Figure 4(b) represents the magnetic configuration of a head-to-head vortex DW in the free layer in such a stray field (superposed on a 60 Oe uniform field). The line represents the "center" of the DW (cells with no longitudinal magnetization). Figure 4(c) compares the morphology of head-to-head and tail-to-tail DW in equivalent fields. The drastic difference between the two kinds of DW might indeed explain the observed difference in the depinning field statistics. Determinations of the depinning field in this idealized geometry do not reproduce quantitavely the measured depinning field. Nevertheless the predicted differences between the P→AP and AP→P transitions (respectively tail-to-tail and head-to-head domain wall) depinning field are 37 Oe for a vortex domain wall and 47 Oe for a tranverse domain wall [23]. These differences are of the same order of magnitude as the experimental ones, confirming the importance of the dipolar field originating from the pinned layer on the depinning process.

To summarize, we have shown the stochastic nature of the injection and depinning fields in a submicron wire structure and we have found that the stray field originating from the pinned layer affects dramatically the distributions of injection and depinning fields. This modification of the "pinning potential" must be considered in quantitative studies of DW depinning by a magnetic field or a spin polarized current. Furthermore, it might also be used to intentionally shape the pinning potential without changing either the structure or the current distribution.

**Figure captions :**

Figure 1 : (a) SEM image of a device. The white arrow indicates the position of the notch. (b) and (c) MFM images of the domain wall trapped respectively in the pad before injection and at the notch vicinity before depinning.

Figure 2 : Normalized resistance recorded during 4 different hysteresis loops.

Figure 3 (color online): (a) Statistical distribution of the injection field for descending field sweep (antiparallel to parallel) in dark blue and ascending field sweep (parallel to antiparallel) in orange. For the late case, absolute value of the field is considered. (b) Same distributions for the depinning field. In inset: set of magnetoresistance measurements.

Figure 4 (color on line): (a) Amplitude of the stray field originating from the pinned layer averaged on the free layer thickness. (b) Magnetic configuration for a head-to-head dowain wall (see text). (c) Comparison between the position of the head-to-head and tail-to-tail DW in equivalent fields.

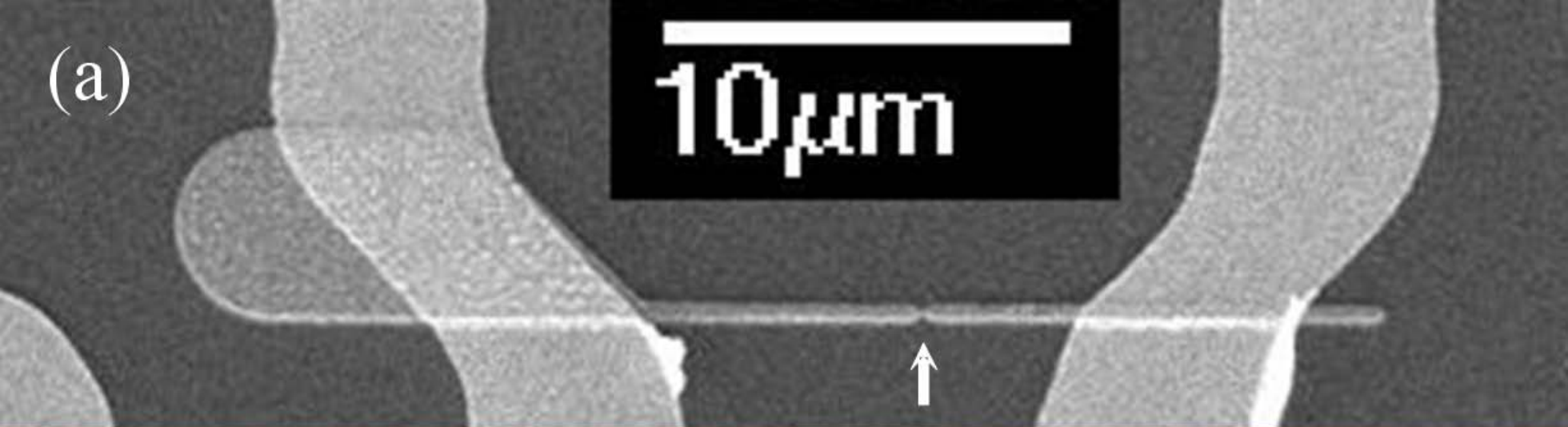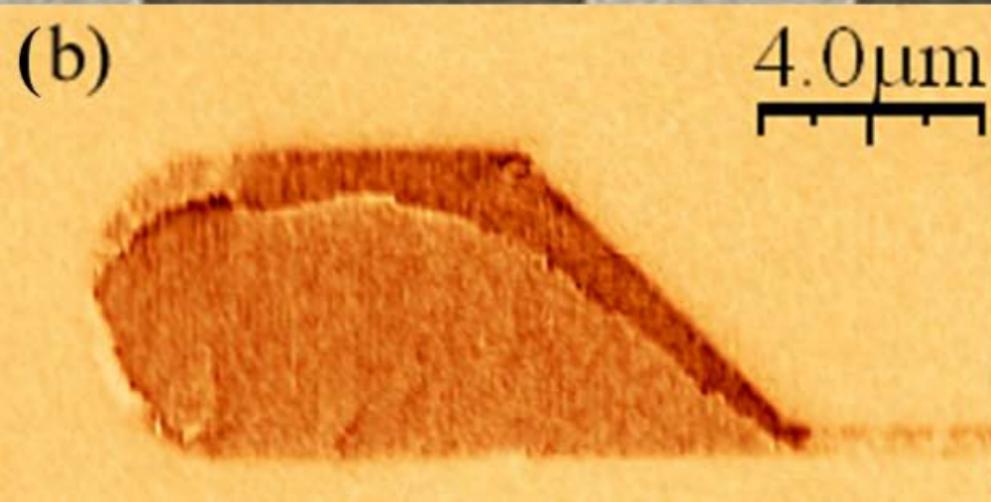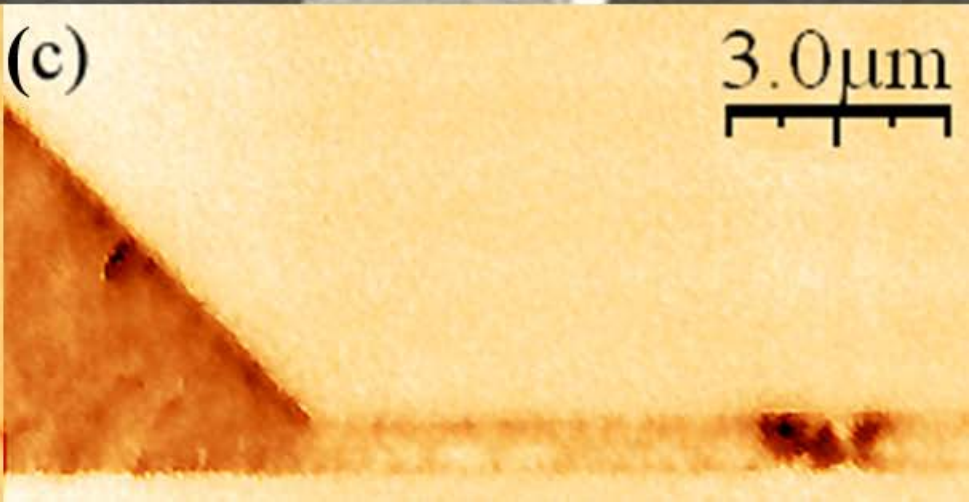

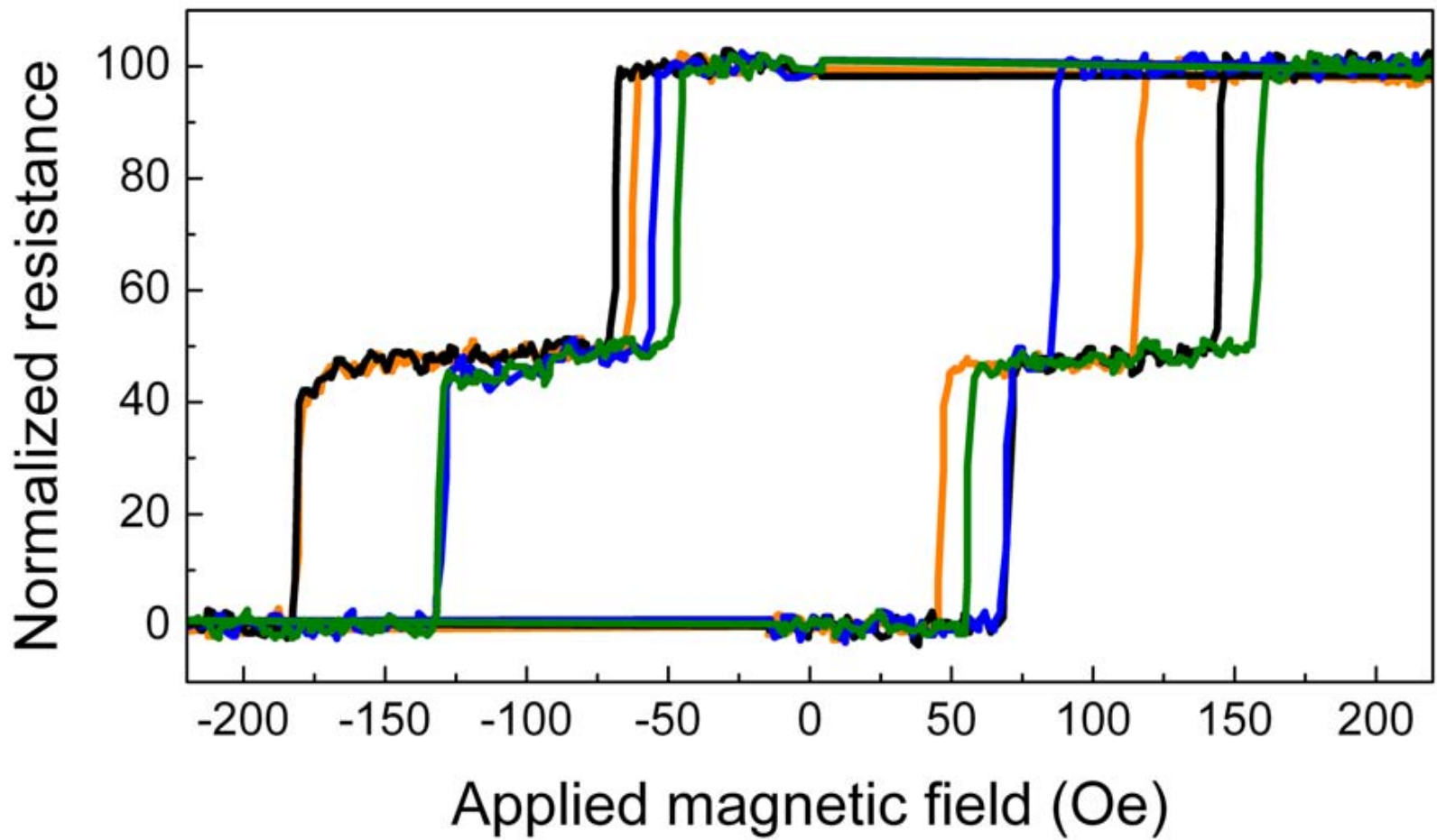

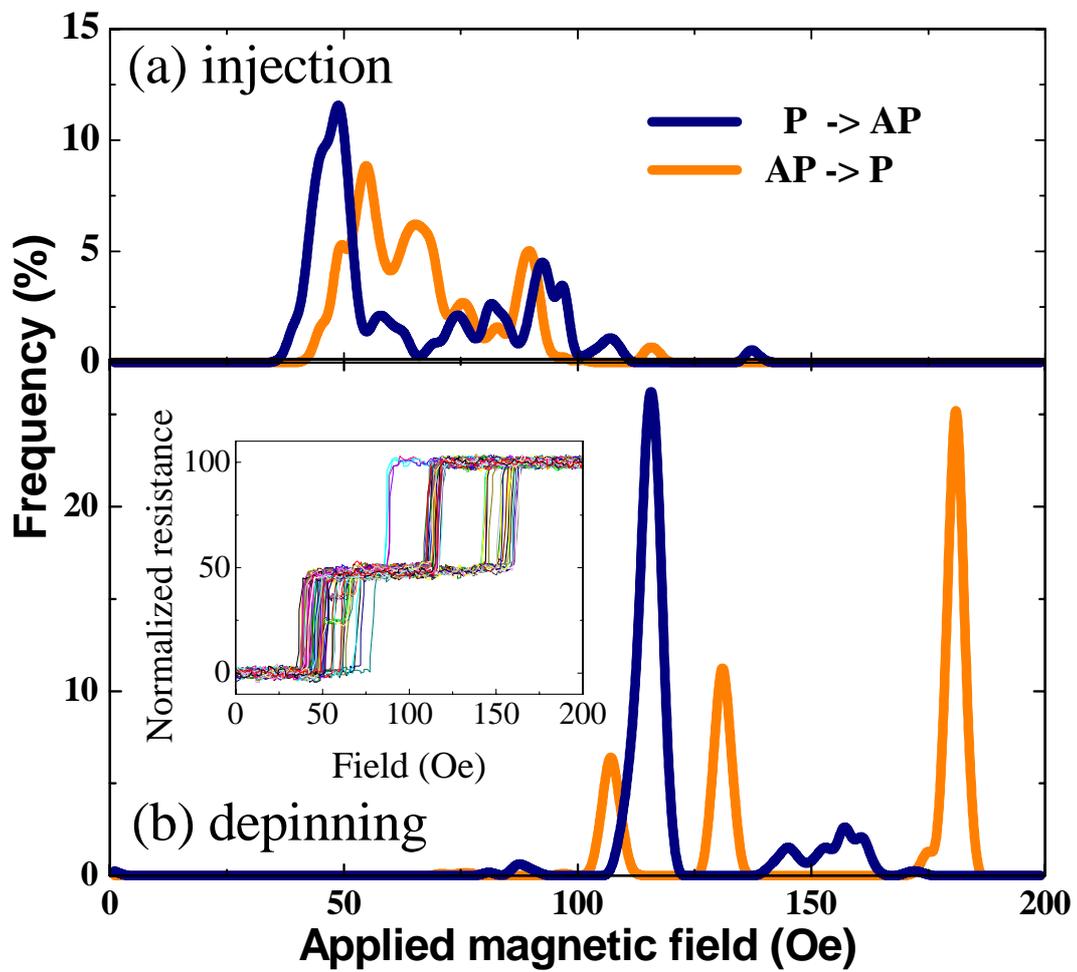

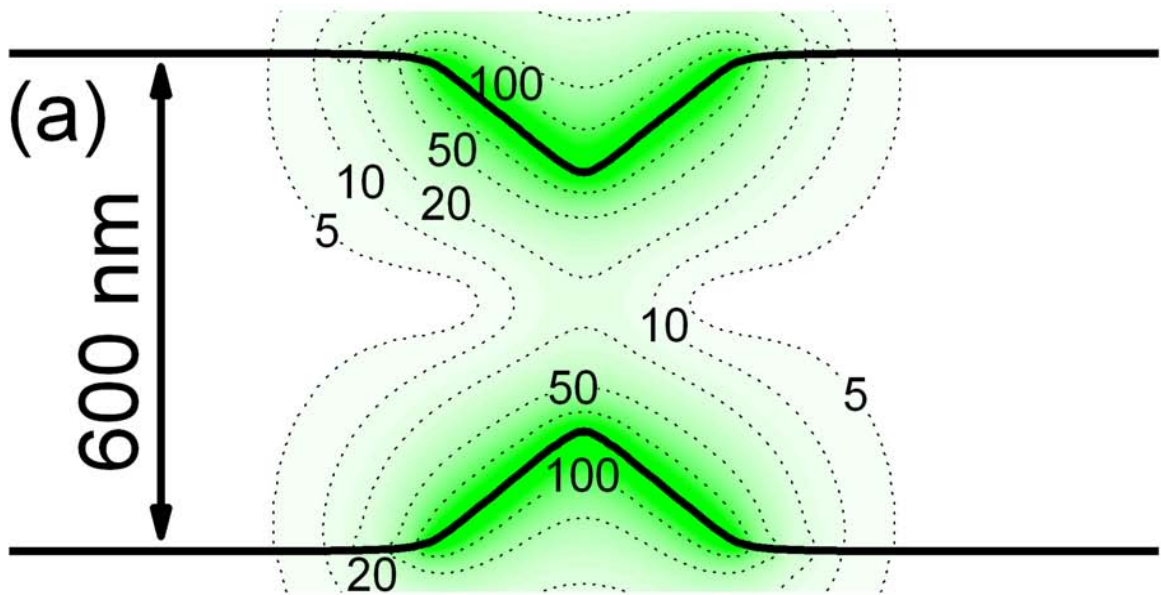
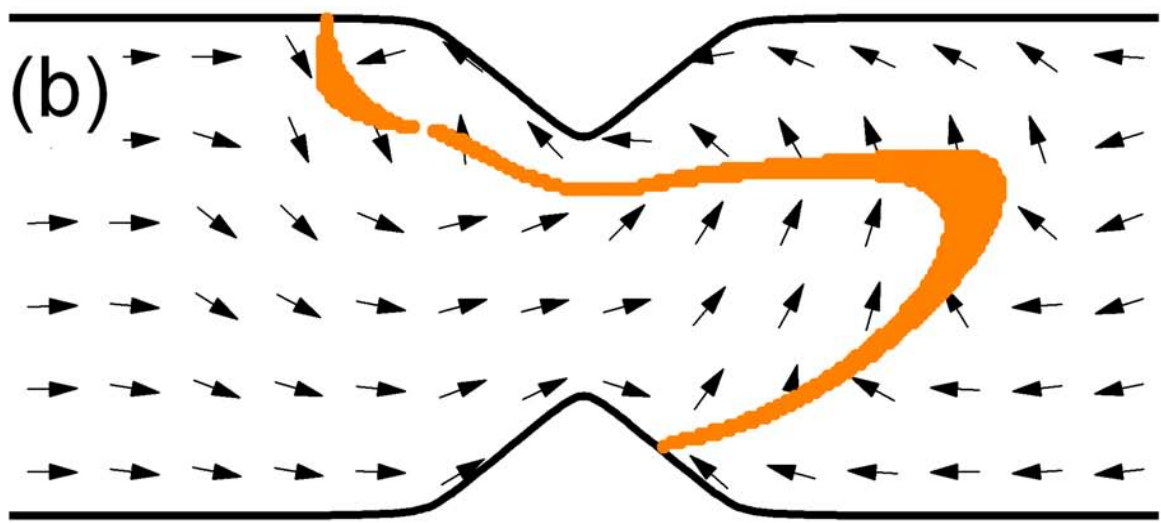
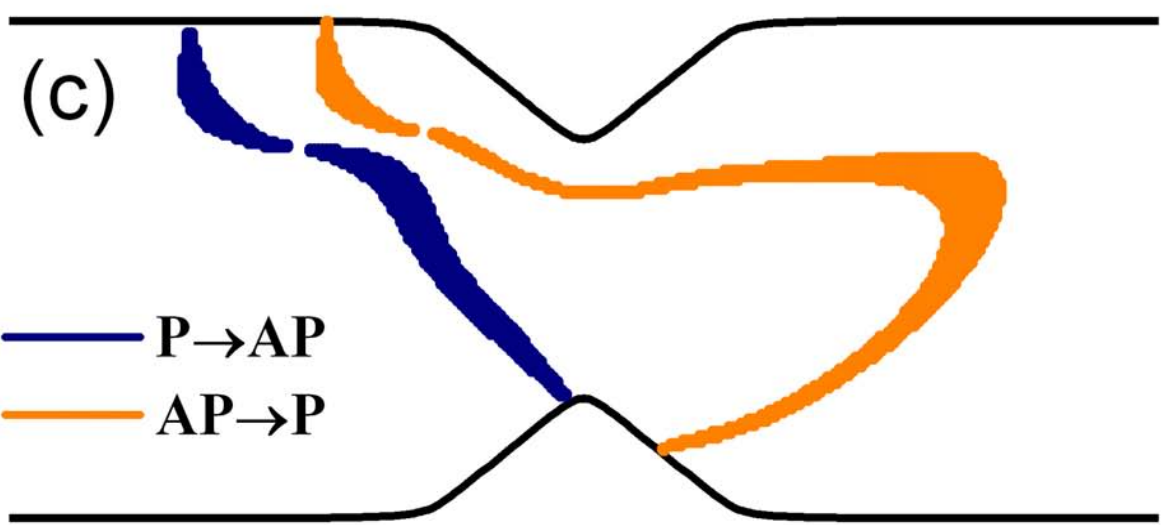